%% file: main.tex
\RequirePackage{fix-cm}
\documentclass[smallextended]{svjour3}       
\smartqed  
\usepackage{graphicx}
\usepackage{graphicx}
\usepackage{amsmath}
\usepackage{amsfonts}
\usepackage{amssymb}
\usepackage{algorithm}
\usepackage{algorithmic}
\usepackage{subfig}
\usepackage{bbold}
\usepackage{hyperref}
\usepackage{nomencl}
\usepackage[toc,page]{appendix}
\usepackage{fontawesome}
\usepackage{hyperref}

\renewcommand{\nompreamble}{The following list defines several notations that will be used throughout the body of the document.}

\journalname{Journal of Fourier Analysis and Applications}

\def\itip{\ref@jnl{IEEE Transactions on Image Processing}} 
\def\physrep{\ref@jnl{Phys.~Rep.}}   

\DeclareUnicodeCharacter{2212}{-}
 
\begin{document}

\title{Galaxy Image Restoration with Shape Constraint
}


\author{Fadi Nammour                \and
        Morgan A. Schmitz           \and
        Fred Maurice Ngol\`e Mboula \and
        Jean-Luc Starck             \and
        Julien N. Girard 
}


\institute{Fadi Nammour \at
              AIM, CEA, CNRS, Universit\'e Paris-Saclay, Universit\'e de Paris, Sorbonne Paris Cit\'e, F-91191 Gif-sur-Yvette, France \\
              \email{fadinammour95@gmail.com}           
           \and
           Morgan A. Schmitz \at
              Department of Astrophysical Sciences, Princeton University, 4 Ivy Ln., Princeton, NJ08544, USA \\
              \email{morgan.schmitz@astro.princeton.edu}           
            \and
           Fred Maurice Ngol\`e Mboula \at
              Institut LIST, CEA, Universit\'e Paris-Saclay, F-91191 Gif-sur-Yvette, France \\
              \email{fred-maurice.ngole-mboula@cea.fr}           
            \and
           Jean-Luc Starck \at
              AIM, CEA, CNRS, Universit\'e Paris-Saclay, Universit\'e Paris Diderot, Sorbonne Paris Cit\'e, F-91191 Gif-sur-Yvette, France \\
              Tel.: +33 (0)1 69 08 57 64\\
              \email{jean-luc.starck@cea.fr}           
            \and
           Julien N. Girard \at
            AIM, CEA, CNRS, Universit\'{e} de Paris, Universit\'{e} Paris-Saclay, F-91191 Gif-sur-Yvette, France \\
            ORCID: 0000-0003-0432-403X\\
              \email{julien.girard@cea.fr}           
}

\date{Received: date / Accepted: date}

\maketitle

\begin{abstract}
\input{abstract.tex}

\end{abstract}

\makenomenclature
\input{notations.tex}
\printnomenclature

\section{Introduction}
\label{intro}
\input{intro.tex}

\section{Galaxies and Shape Measurement}
\label{sec:1}
\input{sec1.tex}

\section{Sparse Deconvolution}
\label{sec:2}
\input{sec2.tex}

\section{Deconvolution with Shape Constraint}
\label{sec:3}
\input{sec3.tex}

\section{Numerical Experiments}
\label{sec:4}
\input{sec4.tex}

\section{Reproducible Research}
\label{sec:5}
\input{sec5.tex}

\section{Conclusion}
\label{conclu}
\input{conclu.tex}

\begin{acknowledgements}
\label{ack}
\input{ack.tex}
\end{acknowledgements}

\clearpage

%
%

\begin{appendices}
\section{Expressing Galaxy Ellipticity with Inner Products}
\label{append:ell2inner_prod}
\input{append_A.tex}

\section{Dataset Generation}
\label{append:dataset}
\input{append_B.tex}

\end{appendices}
\clearpage

\bibliographystyle{spmpsci}      
\bibliography{references}   

\end{document}

%% file: abstract.tex

Images acquired with a telescope are blurred and corrupted by noise. 
The blurring is usually modeled by a convolution with the Point Spread Function and the noise by Additive Gaussian Noise. 
Recovering the observed image is an ill-posed inverse problem. Sparse deconvolution is well known to be an efficient deconvolution technique, 
leading to optimized pixel Mean Square Errors, but without any guarantee that the shapes of objects (e.g. galaxy images) contained in the data will be preserved. 
In this paper, we introduce a new shape constraint and exhibit its properties. By combining it with a standard  sparse regularization in
the wavelet domain, we introduce the Shape COnstraint REstoration algorithm (SCORE),  which performs a standard sparse deconvolution, while 
preserving galaxy shapes. We show through numerical experiments that this new approach leads to a reduction of galaxy ellipticity measurement errors by at least 44\%. \href{https://github.com/CosmoStat/score}{\faGithub}

%% file: notations.tex
\nomenclature[01]{$\ast$}{Convolution operator}
\nomenclature[02]{$\odot$}{Element-wise multiplication operator}
\nomenclature[03]{$"\"|\cdot"\"|_1$}{$\ell_1$-norm}
\nomenclature[04]{$"\"|\cdot"\"|_2$}{Euclidean norm}
\nomenclature[05]{$"\"|\cdot"\"|_\text{F}$}{Frobenius norm}
\nomenclature[12]{$\iota_+$}{Moreau's indicator function of the vector set with non-negative entries}
\nomenclature[07]{$v[k]$}{The $k^{\text{th}}$ element of the vector $v$.}
\nomenclature[08]{$v_\pi$}{The rotation by $v$ of $\pi$ radians, i.e. $\forall k \in \{1,\dots,n^2\} \, , \, v_\pi\left[k\right] = v\left[n^2+1-k\right]$ with $v  \in \mathbb{R}^{n^2}$, vector representation of an $\mathbb{R}^{n\times n}$ image }
\nomenclature[09]{$\rho$}{Function that returns the spectral radius of a matrix}
\nomenclature[10]{$I_n$}{Identity matrix of size n}
\nomenclature[11]{$\mathbf{1}_n$}{All-ones matrix of size n}
\nomenclature[06]{$\left<\cdot,\cdot\right>$}{Canonical inner product}
\nomenclature[13]{sgn}{Sign function}

%% file: intro.tex
 Every  acquisition system generates images with imperfections. Generally, the structure of the system induces a blurring of the images. This blur is often modeled using a Point Spread Function (PSF). Here, we will consider this PSF to be space-invariant and denote it $h \in \mathbb{R}^{n\times n}$. In addition, the sensors' variations are likely to introduce noise in the image. We consider this noise to be additive and denote it $b \in \mathbb{R}^{n\times n}$. The observational model is then as follows:
\begin{equation} \label{eq:invprob}
    y = x_T \ast h + b \quad,
\end{equation}
where $x_T \in \mathbb{R}^{n \times n}$ is the ground truth image, and $y\in\mathbb{R}^{n\times n}$ is the observed image. We can partially restore $y$ by applying the least squares method. In this case, the solution is oscillates 
because the problem in eq.~\ref{eq:invprob} is ill-conditioned. An exploratory work~\cite{shakibaei2014image} has been done for image deconvolution in the moments space, however the PSF profile was assumed to be Gaussian and the observed image did not contain noise. More generally, it is an ill-posed problem and can, instead, be tackled using regularization~\cite{bertero1998introduction}. For this purpose, we can add constraints related to the signal's energy, its derivatives~\cite{bertero1998introduction}, such as total variation~\cite{rudin1992nonlinear,chambolle2010introduction}, or its sparsity~\cite{starck2015sparse,farrens2017space}, where sparsity measures the number of non-zero elements in a signal. These commonly used methods are well suited for solutions that optimize the Mean Square Error (MSE). The MSE of an estimation, $x\in\mathbb{R}^{n\times n}$, of $x_T$, is
\begin{equation}\label{eq:MSE}
    \text{MSE}(x)=\sum_{i=1}^m\left(x_T[i]-x[i]\right)^2 \quad.
\end{equation}
While standard, the MSE is not always the target criterion we want to optimize. For instance, in astrophysics, the \textit{shape} of galaxies (often encoded through a measure of its ellipticity) is central to many scientific goals, such as in weak gravitational lensing~\cite{mandelbaum2018} or galaxy evolution studies~\cite{rodriguez2016}. There is, however, no guarantee that the deconvolution process, especially if non-linear, preserves the  galaxies' shapes. 
For this reason, full galaxy image deconvolution is rarely used in practice in weak lensing studies. The effect of the PSF is, instead, accounted for using either moments-based methods, such as the Kaiser-Squire-Broadhurst (KSB) approach~\cite{kaiser1994method}, or forward modeling, assuming an analytical profile for the galaxy~\cite{miller2007}. Note neither of these approaches produces a deconvolved image of the original galaxy. If such a profile is required to achieve some other science goals, an explicit deconvolution would be performed separately and the resulting image would not possess the correct ellipticity. In radio-interferometry, the situation is even more problematic since data are acquired in Fourier space and these standard ellipticity measurement techniques require first reconstructing an image. In \cite{patel2014,patel2016}, it was shown that standard radio-interferometry image reconstruction techniques could not be used to obtain 
reliable measurements, which led the community to develop fitting techniques  in Fourier space \cite{rivi2016,rivi2018}. 
Since each Fourier component contains information about all galaxies, it therefore requires simultaneously fitting  the ellipticities of all galaxies contained in 
the observed image. Such a minimization is rather complex and relies on the use of time-consuming Hamiltonian Monte Carlo techniques.
An alternative and original approach was proposed in \cite{kumar2017deblurring,shakibaei2014image}, using
a relationship between the moments of the degraded image and the moments of the original image and the PSF. The solution is then obtained by inverting the moment equation.
This method, however, relies on the point spread function being an elliptical Gaussian, which is not the case in practice. 

In this paper, we propose an intermediate solution, between the geometric moment method and a standard regularized deconvolution technique such as sparsity, by embedding a shape constraint derived from the moments in a restoration framework. 

In sect.~\ref{sec:1} and~\ref{sec:2}, we start by formulating the constraint using the analytical expression of the ellipticity. 
In sect.~\ref{sec:3}, we exhibit its main properties, which will allow us to build the Shape COnstraint REstoration algorithm (SCORE). 
SCORE is a sparse restoration algorithm, into which the proposed constraint is plugged, leading  to the first shape constraint deconvolution algorithm.
 Finally, in sect.~\ref{sec:4}, we present the results of numerical experiments.

%% file: sec1.tex

The definition of ellipticity is straightforward for an object whose light profile has elliptical isophotes. In order to generalize this definition to objects with arbitrary profiles, the statistical moments of its light profile (or observed image, in practice) are used. For a galaxy image, $x\in \mathbb{R}^{n \times n}$, let us define its complex ellipticity $e(x) = e_1(x) + \bold{i}e_2(x)$, as~\cite{bartelmann2001}
\begin{equation} \label{eq:ell_mom}
    \mathrm{e}_1(x) = \frac{\mu_{2,0}(x)-\mu_{0,2}(x)}{\mu_{2,0}(x)+\mu_{0,2}(x)}
    \quad \mathrm{and} \quad
    \mathrm{e}_2(x) = \frac{2\mu_{1,1}(x)}{\mu_{2,0}(x)+\mu_{0,2}(x)}
    \quad,
\end{equation}
\
where $\mu_{s,t}(x)$ are the centered moments. In the case of a discrete image containing an object of interest, an unbiased estimator for these can be computed as follows:
\begin{equation} \label{eq:centered_mom}
    \mu_{s,t}(x)=\sum_{i=1}^n\sum_{j=1}^n x[(i-1) n+j] (i-i_c)^s (j-j_c)^t \quad,
\end{equation}
where $i_c = \frac{\sum_{i=1}^n\sum_{j=1}^n i\cdot x[(i-1) n+j]}{\sum_{i=1}^n\sum_{j=1}^n x[(i-1) n+j]}$ and $j_c = \frac{\sum_{i=1}^n\sum_{j=1}^n j\cdot x[(i-1) n+j]}{\sum_{i=1}^n\sum_{j=1}^n x[(i-1) n+j]}$ are the coordinates of the centroid of $x$.\\
\textbf{Remark: }In eq.~\ref{eq:centered_mom}, $i$ and $j$ correspond respectively to the $i^{\text{th}}$ row and $j^{\text{th}}$ column of the galaxy image, $x$.

\begin{figure}
  \begin{tabular}{cc} 
    \includegraphics[width=0.45\textwidth]{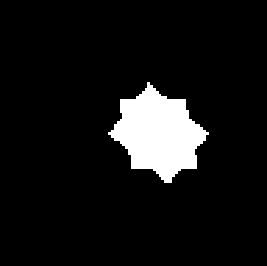}
    &
    \includegraphics[width=0.45\textwidth]{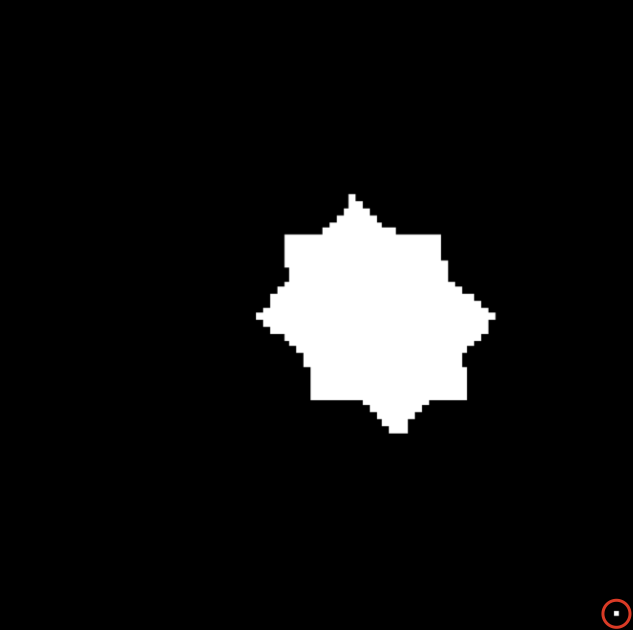}
  \end{tabular}
\caption{These two binary images only differ by a single pixel, encircled in red on the bottom right of the image on the right. The estimated ellipticity, using \eqref{eq:ell_mom}, is $0.0016+\mathbf{i}0.2196$ for the left image and $-0.0094+\mathbf{i}0.1916$ for the right one. This corresponds to a deviation of 14\%. \href{https://github.com/CosmoStat/score/blob/master/reproducible_research/paper_results/ellipticity_sensitivity_example.ipynb}{\faFileCodeO}}
\label{fig:EllDiffIm}       
\end{figure}

These quantities are extremely sensitive to noise. The practical computation of galaxy ellipticities is thus an ill-conditioned problem, as illustrated in Fig.~\ref{fig:EllDiffIm}. A common way to add robustness, i.e. to reduce sensitivity to background noise, is through the use of a window function, $g\in\mathbb{R}^{n\times n}$, typically chosen to be a 2-dimensional Gaussian. The size of this window function is either fixed a priori, or fitted to $y$, the observed image~\cite{hirata2003}. An estimator of the ellipticity of $y$, noted $\mathrm{e}_\text{int}(y) = \mathrm{e}_{\text{int},1}(y)+\mathbf{i}\mathrm{e}_{\text{int},2}(y)$, is then
\begin{equation} \label{eq:ell_int}
    \mathrm{e}_{\text{int}}(y) = \frac{\mu_{2,0}(y\odot g)-\mu_{0,2}(y\odot g)+2\bold{i}\mu_{1,1}(y\odot g)}{\mu_{2,0}(y\odot g)+\mu_{0,2}(y\odot g)}\quad.
\end{equation}

The classical KSB method~\cite{kaiser1994method} and its later improvements~\cite{viola2011biases} are based on such estimators. The ellipticity of a weighted version of the observed image is first computed in order to reduce noise effects. Then the PSF effects are corrected in the moments space. The keystone of this method is approximating the PSF effects linearly under the assumption that the PSF has slight anisotropies. In the following, we will show that a full restoration of an image is possible, preserving the galaxy shape information.

%% file: sec2.tex
\label{subsec:2_1}

Sparsity has proven to be an effective regularization technique for denoising~\cite{farrens2017space,starck2015sparse}. Its use in deconvolution along with positivity offers satisfying results regarding the pixel error~\cite{farrens2017space}. Sparse regularization is applied in a space where the solution is known to be sparse. In the case of galaxy images, it has been shown in~\cite{starck2015sparse} that starlets offer such a sparse representation.

In the following, we will assume that the noise, $b$ in eq.~\ref{eq:invprob}, is white additive Gaussian noise with variance $\sigma^2I_n$.

Let $\phi=\left(\phi_i\right)_{i\in\left\{1,\dots,I \right\}}$ denote the starlet transform operator, with $I$ its chosen number of components. The loss function for the sparse deconvolution problem can be written as the sum of differentiable and non-differentiable terms:
\begin{equation}\label{eq:L0}
    L_0(x) = \overbrace{\underbrace{\frac{1}{2\sigma^2}\|x\ast h-y\|^2_2}_{\text{data-fidelity}}}^{:=L_{0d}(x),\text{ differentiable}}+\overbrace{\underbrace{\|\lambda_0 \odot \phi(x)\|_1}_{\text{sparsity}}+\underbrace{\iota_+(x)}_{\text{positivity}}}^{:=L_{0p}(x),\text{ non-diffrentiable}} \quad,
\end{equation}
where $\lambda_0$ is a weighting matrix with non-negative entries.

We now give the major properties of $L_0$ needed to construct a Sparse Restoration Algorithm (SRA) that minimizes it. Straightforwardly, $L_{0d}$ has gradient
\begin{equation}
    \nabla L_{0d}(x) = \frac{1}{2\sigma^2}2h_\pi \ast \left(x\ast h -y\right) \quad.
\end{equation}
Following from its definition, the Lipschitz constant of $\nabla L_{0d}$, noted $\alpha_0$, is
\begin{equation}
    \alpha_0 = \frac{1}{2\sigma^2}\rho\left(2h_\pi\ast h \ast I_{n^2}\right) \quad.
\end{equation}

Following \cite{starck2015sparse}, we set the value of $\lambda_0$ such that it is proportional to the standard deviation map of $\phi \nabla L_{0d}(x_T)$. We notice that for $x=x_T$, we have $x\ast h - y$ equal to $-b$ which is also a white Gaussian noise of variance $\sigma^2I_{n}$. Consequently,
\begin{equation}
    \phi_i \ast \nabla L_{0d}(x_T) = -\frac{1}{\sigma^2}\phi_i \ast h_\pi\ast b \quad, \forall i \in \{1,\dots,I\}\,.
\end{equation}
It follows that $\phi \nabla L_{0d}(x_T)$ is colored Gaussian noise, with variance
\begin{equation}
    \Sigma_0 = \underbrace{\frac{1}{\sigma^2}\left[\left(\phi_i \ast h_\pi\ast I_{n^2}\right)\left(\phi_i \ast h_\pi\ast I_{n^2}\right)^\top\right]_{i\in\{1,\dots,I\}}}_{{:=\left(\Sigma_{0i}\right)}_{i\in\{1,\dots,I\}}} \quad.
\end{equation}
We then set
\begin{equation}\label{eq:lambda0}
    \lambda_0 = \left[\kappa[i] \cdot \text{diag}\left(\Sigma_{0i}\right)\right]_{i\in\{1,\dots,I\}}=\frac{1}{\sigma^2}\left(\kappa[i] \cdot \|\phi_i \ast h_\pi\|^2_2\mathbf{1}_{n^2}\right)_{i\in\{1,\dots,I\}}\quad,
\end{equation}
where $\kappa$ is a vector in $\mathbb{R}^{I+1}$ of the form $\left(0,q,\dots,q,q+1\right)$, assuming that the components of $\phi$ are arranged gradually from the coarse scale $\phi_0$ to the finest scale $\phi_I$. In this work, we set $q=4$.

Finally, we approximate the proximal operator of the non-differentiable part, $L_{0p}$, of the  the loss function. To do so, let us first recall the exact forms of the proximal operators of $\iota_+$ and $\|\lambda \odot\cdot\|_1$, with $\lambda\in\mathbb{R}^{(I+1)\times n \times n}_+$:
\begin{equation}
    \text{prox}_{\iota_+}(x) = (x)_+\quad, 
\end{equation}
where $\forall k \in\{1,\dots,n^2\}$,
\begin{equation}
(x)_+[k] = \text{max}\left(x[k],0\right)\quad,
\end{equation}
and
\begin{equation}
    \text{prox}_{\|\lambda \odot.\|_1}(x) = \text{ST}_\lambda (x)\quad,
\end{equation}
where $\text{ST}_\lambda$ is the soft-thresholding operator, defined $\forall k \in\{1,\dots,(I+1)n^2\}$ as 
\begin{equation}
    \text{ST}_\lambda(x) [k]=\left\{\begin{array}{ll}
         x[k]-\text{sgn}\left(x[k]\right)\lambda \text{ if }\left|x[k]\right|\geq \left|\lambda[k]\right|\\
         0 \text{ otherwise}
    \end{array} \right.\quad.
\end{equation}
Nevertheless, in practice, the hard-thresholding operator is preferred over the soft-thresholding in order to reduce the bias introduced by image restoration~\cite{starck2002nonlinear,blumensath2009iterative}. Let $\text{HT}_\lambda$ denote the hard-thresholding operator, defined $\forall k \in\{1,\dots,(I+1)n^2\}$ as
\begin{equation}
    \text{HT}_\lambda(x) [k]=\left\{\begin{array}{ll}
         x[k] \text{ if }\left|x[k]\right|\geq \left|\lambda[k]\right|\\
         0 \text{ otherwise}
    \end{array} \right.\quad.
\end{equation}
From these, we define
\begin{equation}
    \text{p}_{\lambda_0}(x) = \left[\phi^{-1}\left(\text{HT}_{\lambda_0}\left[\phi\left(x\right)\right]\right)\right]_+\quad,
\end{equation}
the approximation of $\text{prox}_{L_{0p}}$ we will use in the present work.
\\
\textbf{Remark: }$\text{p}_{\lambda_0}$ relies on two approximations. The first is related to the starlet, which is redundant (thus non-orthogonal), as an orthogonal transform. The second is due to assuming that the proximal operator of the sum of two non-differentiable terms is the composition of the proximal operators of the terms, which does not hold in general.

The implementation of SRA used in this work is based on a proximal splitting frame, more precisely, on forward-backward splitting methods~\cite{starck2015sparse}. The resulting algorithm is given in alg.~\ref{alg:sra}. We still need to set the stopping criterion, $A$; the first-guess, $t \in \mathbb{R}^{n\times n}$; $\alpha_{\epsilon}$ and $\hat{\lambda}_0$ which are respectively the estimations of $\alpha_0$ and $\lambda_0$. 

\begin{algorithm}
\caption{SRA algorithm}
\label{alg:sra}
\begin{algorithmic}
\STATE{\textbf{Task:} Restore $x_T$ using $y$ and $h$.}
\STATE{\textbf{Parameters:} $\gamma$, $\epsilon >0$, boolean $A$.}
\STATE{\textbf{Initialization: $x^{(0)}\gets t$, $\beta \gets \alpha_{\epsilon}^{-1}$}}
\WHILE{not$\left(A\right)$}
\STATE{$x^{(i+1)}\gets \text{p}_{\beta \hat{\lambda}}\left[x^{(i)}-\beta \nabla L_d \left(x^{(i)}\right)\right]$
\STATE{$i\gets i+1$}
}
\ENDWHILE
\RETURN $x^{(i)}$
\end{algorithmic}
\end{algorithm}

We compute $\alpha_\epsilon$ by using the power iteration method to obtain an estimation of $\alpha$, and then we multiply the output by $(1+\epsilon)$ to make sure that we did not go below the lowest upper bound (for this paper, we set $\epsilon=0.05$). $\hat{\lambda}_0$ is computed using eq.~\ref{eq:lambda0}. And we set $t$ to $\frac{1}{n^2}\mathbf{1}_{n^2}$. 

For the stopping criterion, we considered two cases :
\begin{description}
    \item[\textbf{The denoising case} ($h=\delta$)\textbf{:}]
    here the problem is well-conditioned, and we set $A$ to `$i\leq N_i$' where $N_i$ is the number of iterations. For the numerical experiments we set $N_i$ to 40.
    
    \item[\textbf{The general case:}] here the problem is ill-conditioned, prompting us to set $A$ to `$i\leq N_i$ and $\left|\frac{\left[L\left(x^{(i)}\right)+L\left(x^{(i-1)}\right)\right]-\left[L\left(x^{(i-2)}\right)+L\left(x^{(i-3)}\right)\right]}{L\left(x^{(i-2)}\right)+L\left(x^{(i-3)}\right)}\right|\leq c$'. In the present experiments, we set $N_i=150$ and $c=10^{-6}$.
\end{description}

%% file: sec3.tex
\subsection{The Shape Constraint}
\label{subsec:2_2}

Ideally, the shape constraint should take the form of a data fidelity term in a space that corresponds to the ellipticity. However the ellipticity, $e(x)$, as defined in eq.~\ref{eq:ell_mom} is a non-linear function of the galaxy image, $x$. We thus express it as a combination of linear quantities, which will prove easier to handle mathematically. In~\cite{bernstein2014bayesian}, it is shown that the ellipticity can be rewritten using scalar products. Analogously, we derive in appendix~\ref{append:ell2inner_prod} the following formulae:
\begin{equation} \label{eq:ell_scal}
    \begin{aligned}
    &\mathrm{e}_1(x) = \frac{\left<x,u_3\right>\left<x,u_5\right>-\left<x,u_1\right>^2+\left<x,u_2\right>^2}{\left<x,u_3\right>\left<x,u_4\right>-\left<x,u_1\right>^2-\left<x,u_2\right>^2} \quad,
    \\
    &\mathrm{e}_2(x) = \frac{2(\left<x,u_3\right>\left<x,u_6\right>-\left<x,u_1\right>\left<x,u_2\right>)}{\left<x,u_3\right>\left<x,u_4\right>-\left<x,u_1\right>^2-\left<x,u_2\right>^2}
    \quad,
    \end{aligned}
\end{equation}
with $\left(u_k\right)_{k\in\{1,\dots,6\}}$ in $\mathbb{R}^{6\times n\times n}$, defined for all $i$ and $j$ in $\{1,\dots,n\}$ as
\begin{equation} \label{eq:u_i}
    \begin{aligned}
        &u_1[(i-1)n+j] = (i), &&u_2[(i-1)n+j] = (j),\\
        &u_3[(i-1)n+j] = (1), &&u_4[(i-1)n+j] = (i^2+j^2),\\
        &u_5[(i-1)n+j] = (i^2-j^2), &&u_6[(i-1)n+j] = (ij).
    \end{aligned}
\end{equation}

Eq.~\ref{eq:ell_scal} shows that the ellipticity information is contained in the set of $6$ scalar products. Additionally, in eq.~\ref{eq:u_i}, we can see that $\left(u_k\right)_{k\in\{1,\dots,6\}}$ are constant vectors. The scalar products are therefore all linear functions of $x$. Consequently, we choose them as building blocks for the shape constraint, instead of directly using the ellipticity (which is not a linear function of $x$). From this, we give a preliminary formulation of the constraint:
\begin{equation}\label{eq:m0}
    M_0(x) = \sum_{i=1}^6 \omega_i \left<x\ast h-y,u_i\right>^2 \quad,
\end{equation}
where the components of $\left(\omega_i\right)_{i\in\{1,\dots,6\}}$ are real valued scalar weights.\\

As discussed in  sect.~\ref{sec:1}, the quantities in eq.~\ref{eq:m0} are extremely sensitive to noise. A natural way to increase the robustness of our shape constraint would be, in analogy with the ellipticity estimator of eq.~\ref{eq:ell_int}, to apply a weighting function $g$. This choice, however, comes with the burden of correctly choosing $g$. A fixed window function would lack flexibility and likely lead to poor estimators of ellipticity for some objects $y$. Fitting $g$ to $y$ would improve flexibility, but require an additional preprocessing step. 

\begin{figure}
  \includegraphics[width=\textwidth]{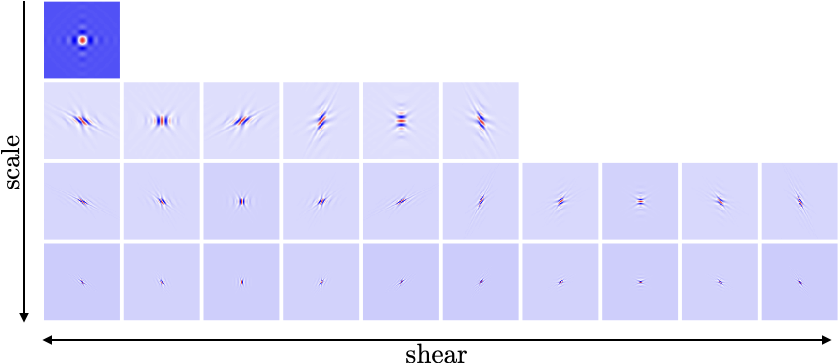}
\caption{Representation of the shearlet bands with 3 scales.}
\label{fig:shearlets}       
\end{figure}

An alternative approach is to apply the constraint on many windows of different sizes and orientations, so that at least one of them is a good fit to $y$. To find such a set of windows, we consider curvelet-like decompositions \cite{starck:sta01_3,starck2015sparse} where all the bands correspond to windows with different orientations, and every scale corresponds to a different size. \textit{Shearlets} are particularly appropriate for our ends, as can be seen from Fig.~\ref{fig:shearlets}. This choice was also motivated by the two following properties~\cite{kutyniok2012introduction,voigtlaender2017analysis}:
\begin{itemize}
    \item{\textbf{Anistropy}:} Ellipticity is, itself, a measure of anisotropy and the use of the shearlets, which are an anisotropic transform, should help us discriminate objects according to this criterion.
    \item{\textbf{Grid conservation:}} The scaling and shearing 
    operations that transition from one shearlet band to another conserve the points on the grid, which adds numerical stability.
\end{itemize}

Let $\psi=\left(\psi_j\right)_{j \in\{1,\dots,J\}}$ denote the shearlet transform operator, with $J$ its chosen number of components. We formulate the shape constraint as follows:
\begin{equation}
 M(x)=\sum_{i=1}^6\sum_{j=1}^J\omega_{ij}\left<\psi_j(x\ast h)-\psi_j(y),u_i\right>^2\quad,
\end{equation}
where $\left(\omega_{ij}\right)_{\substack{i\in \{ 1,\dots,6 \}\\ j\in \{ 1,\dots,J \}}}$ are real-valued scalars (see sect.~\ref{subsec:3_2} for their practical selection).
By taking into account the fact that the shearlet transform is a linear operator, and denoting $\psi_j^\ast$ the adjoint operator of $\psi_j$ for all $j$ in $\{1,\dots,J\}$, we have
\begin{equation} \label{eq:SC}
    M(x) = \sum_{i=1}^6\sum_{j=1}^J\omega_{ij}\left<x\ast h-y,\psi_j^\ast\left(u_i\right)\right>^2\quad.
\end{equation}

Wavelet moments \cite{chen2011wavelet,farokhi2014near}
and curvelet moments \cite{murtagh2008wavelet,dhahbi2015breast} have been similarly used in the past, but only for classification applications. 

After formulating the constraint, we will put it into use by adding it to a sparse restoration algorithm that computes a solution to eq.~\ref{eq:invprob}. We achieve this by creating the corresponding loss function, exhibiting its properties and, finally, building an algorithm that minimizes it: SCORE.

\subsection{The Loss Function}
\label{subsec:3_1}
Combining the shape constraint from eq.~\ref{eq:SC} and the loss function of SRA from eq.~\ref{eq:L0}, we obtain the SCORE loss function, 
\begin{equation} \label{eq:loss_OSCAR}
    L(x) = \overbrace{\frac{1}{2\sigma^2}\|x\ast h-y\|^2_2+\underbrace{\frac{\gamma}{2\sigma^2}M(x)}_{\text{shape constraint}}}^{:=L_d(x),\text{ differentiable part}}+\overbrace{\|\lambda \odot \phi(x)\|_1+\iota_+(x)}^{:=L_p(x),\text{ non-diffrentiable part}} \quad,
\end{equation}
where $\gamma \in \mathbb{R}_+$ is the trade-off between the data-fidelity term and the shape constraint and $\lambda$ is, as before, a weighting matrix with non-negative entries.

Before expressing the main properties of $L$, let us give a reformulation of its differentiable part, $L_d$. Namely, let us show that it can be recast as a single data-fidelity term with a modified norm. As a starting point, on the one hand, we have
\begin{align}
    M(x) &= \frac{\gamma}{2\sigma^2}\sum_{i,j}\omega_{ij}\left<x\ast h-y,\psi_j^\ast\left(u_i\right)\right>^2 \quad,\\
    &= \frac{\gamma}{2\sigma^2}\sum_{i,j}\omega_{ij}\left(\left[\psi_j^\ast\left(u_i\right)\right]^\top\left[x\ast h-y\right]\right)^\top\left(\left[\psi_j^\ast\left(u_i\right)\right]^\top\left[x\ast h-y\right]\right),\\
    &= \frac{1}{2\sigma^2}\left(x\ast h-y\right)^\top\gamma\underbrace{\sum_{i,j}\overbrace{\omega_{ij}\psi_j^\ast\left(u_i\right)\left[\psi_j^\ast\left(u_i\right)\right]^\top}^{:=Q_{ij} \succeq 0}}_{:=Q \succeq 0}\left(x\ast h-y\right) \quad. \label{eq:sc_vec}
\end{align}
Similarly, on the other hand, we have
\begin{equation}\label{eq:data_fid}
    \frac{1}{2\sigma^2}\|x\ast h-y\|^2_2 = \frac{1}{2\sigma^2}\left(x\ast h -y\right)^\top I_n\left(x\ast h -y\right) \quad.
\end{equation}
By summing eq.~\ref{eq:sc_vec} and~\ref{eq:data_fid}, we obtain 
\begin{align}
    L_d(x) &= \frac{1}{2\sigma^2}\left(x\ast h -y\right)^\top\underbrace{\left( I_n+\gamma Q\right)}_{:=S\succ 0}\left(x\ast h -y\right) \quad,\label{eq:loss_vec}\\
    L_d(x) &= \frac{1}{2\sigma^2}\|x\ast h -y\|_S \quad. \label{eq:loss_norm}
\end{align}
We can thus interpret the weighted data-fidelity term in eq.~\ref{eq:loss_norm} as an extension of the space of the data-fidelity. When using it, we are effectively not only considering the image space by itself, but also taking the space of scalar products of $M$ into account.

\subsection{Properties of $L$}
\label{subsec:3_2}
Analogously to sect.~\ref{subsec:2_1}, we first determine the values of the constants (other than $\gamma$, which we study in detail in sect.~\ref{subsec:4_1}) that appear in $L$, and how to handle its differentiable and non-differentiable parts within an optimization framework.

To determine $\left(\omega_{ij}\right)_{\substack{i\in\{1,\dots,6\}\\j\in\{1,\dots,J\}}}$ in \eqref{eq:SC}, let us impose that the unweighted data-fidelity and the shape constraint exert the same relative influence when $\gamma$ is 1. In addition, without any further prior, we want all components of $Q$ to have equal influence. With no guarantee of orthogonality, we then impose the following conditions
\begin{equation}\label{eq:mu}
    \left\{
    \begin{array}{ll}
        \|I_n\|_\text{F}=\displaystyle \sum_{i=1}^6\sum_{j=1}^J\|Q_{ij}\|_\text{F} \quad,\\
        \|Q_{ij}\|_\text{F}=\|Q_{kl}\|_\text{F}\quad,\,\forall i,k \in \left\{1,\dots,6\right\}\quad,\,\forall j,l \in \left\{1,\dots,J\right\}.
    \end{array}
    \right.
\end{equation}
Solving the system in~\ref{eq:mu}, leads to:
\begin{equation}
    \omega_{ij}=\frac{n}{\left\|\psi_j^*\left(u_i\right)\right\|_2^2} \quad,\,\forall i \in \{1,\dots,6\}\,,\,\forall j \in \{1,\dots,J\}.
\end{equation}
The gradient of $L_d$ follows from eq.~\ref{eq:loss_vec}:
\begin{equation}
    \nabla L_d(x) = \frac{1}{2\sigma^2}2h_\pi\ast S \left(x\ast h -y\right) \quad.
\end{equation}
Its Lipschitz constant, noted $\alpha$, is
\begin{equation}
    \alpha = \frac{1}{2\sigma^2}\rho\left(2h_\pi\ast h\ast S\right) \quad.
\end{equation}
In order to set $\lambda$, we once again propagate residual noise. Since
\begin{equation}
    \phi_i \ast \nabla L_d(x) = -\frac{1}{\sigma^2}\phi_i \ast h_\pi\ast Sb \quad, \forall i \in \{1,\dots,I\}\,,
\end{equation}
$\phi L_d(x)$ is colored Gaussian noise, with variance
\begin{equation}
    \Sigma = \underbrace{\frac{1}{\sigma^2}\left[\left(\phi_i \ast h_\pi\ast S\right)\left(\phi_i \ast h_\pi\ast S\right)^\top\right]_{i\in\{1,\dots,I\}}}_{:=\left(\Sigma_i\right)_{i\in\{1,\dots,I\}}} \quad.
\end{equation}
This allows us to chose
\begin{equation}\label{eq:lambda}
    \lambda = \left[\kappa[i] \cdot \text{diag}\left(\Sigma_i\right)\right]_{i\in\{1,\dots,I\}}\quad.
\end{equation}
Lastly, similar to sect.~\ref{subsec:3_1}, we approximate $\text{prox}_{L_p}$ by
\begin{equation}
    \text{p}_\lambda(x) = \left[\phi^{-1}\left(\text{HT}_\lambda\left[\phi\left(x\right)\right]\right)\right]_+\quad.
\end{equation}

\subsection{Algorithm}
\label{subsec:3_3}

The SCORE algorithm is given in alg.~\ref{alg:score}. As with sect.~\ref{subsec:2_1}, we also need to set the stopping criterion, $A$; the first-guess, $t \in \mathbb{R}^{n\times n}$; $\alpha_\epsilon$ and $\hat{\lambda}$ which are respectively the estimations of $\alpha$ and $\lambda$.

\begin{algorithm}
\caption{SCORE algorithm}
\label{alg:score}
\begin{algorithmic}
\STATE{\textbf{Task:} Restore $x_T$ using $y$ and $h$.}
\STATE{\textbf{Parameters:} $\gamma$, $\epsilon >0$, boolean $A$.}
\STATE{\textbf{Initialization: $x^{(0)}\gets t$, $\beta \gets \alpha_{\epsilon}^{-1}$}}
\WHILE{not$\left(A\right)$}
\STATE{$x^{(i+1)}\gets \text{p}_{\beta \hat{\lambda}}\left[x^{(i)}-\beta \nabla L_d \left(x^{(i)}\right)\right]$
\STATE{$i\gets i+1$}
}
\ENDWHILE
\RETURN $x^{(i)}$
\end{algorithmic}
\end{algorithm}

As for alg.~\ref{alg:sra}, we compute $\alpha_\epsilon$ by using the power iteration method to obtain an estimation of $\alpha$, and then multiplying the output by $(1+\epsilon)$ (for this paper, $\epsilon=0.05$). 
For the other variables, we considered:
\begin{description}
    \item[\textbf{The denoising case} ($h=\delta$)\textbf{:}]
    here the problem is well-conditioned, therefore we set $A$ to `$i\leq N_i$' where $N_i$ is the number of iterations. To set $t$, we gave it the value of the output of SRA. Finally, we directly compute $\hat{\lambda}$ using the formula in eq.~\ref{eq:lambda}.
    
    \item[\textbf{The general case:}] here the problem is ill-conditioned, which leads us to set $A$ to `$i\leq N_i$ and $\left|\frac{\left[L\left(x^{(i)}\right)+L\left(x^{(i-1)}\right)\right]-\left[L\left(x^{(i-2)}\right)+L\left(x^{(i-3)}\right)\right]}{L\left(x^{(i-2)}\right)+L\left(x^{(i-3)}\right)}\right|\leq c$'. We choose the first guess $t=\frac{1}{n^2}\mathbb{1}$. To compute $\hat{\lambda}$, we generate $G$ realisations of white Gaussian noise of variance $\sigma^2I_{n^2}$. In this paper, we set $c$ to $10^{-6}$ and $G$ to 100.
\end{description}

Assuming that each image contains only one galaxy such that all of its active pixels are connected, we add a post-processing step to remove the other isolated blobs in the output image. To do so, we mask the isolated blobs by first binarizing each output image using its 80\textsuperscript{th} percentile pixel value as a threshold. Then, under the safe assumption that the galaxy of interest should correspond to the largest blob, we set every other blob's pixels to 0.

%% file: sec4.tex
In this section, we perform numerical experiments on simulated galaxy images. We start by describing the dataset, then detail the implementation framework used. 
We also present two experiments, one on denoising and one on deconvolution.

\subsection{Dataset \& Implementation}
\label{subsec:4_1}
To build our dataset, we generate 300 galaxy images, simulated using parameters fitted on real galaxies from the catalog COSMOS~\cite{Mandelbaum_2011} and 300 PSF images with a Moffat profile. Each image has $96\times96$ pixels. For further details on the data generation, see appendix ~\ref{append:dataset}. To create the observations, we convolve each galaxy with a PSF then add noise.

Regarding noise levels, we use the following definition for the signal-to-noise ratio ($\text{SNR}$) of an observation $y$ of $x_T$:
\begin{equation*}
    \text{SNR}(y) = \frac{\left\|x_T\right\|_2}{\sigma} \quad.
\end{equation*}
The chosen SNR levels are 40, 75, 150 and 380, with 300 observations generated for each.

The implementation was done using \texttt{Python 3.6.8}, \texttt{ModOpt 1.3.0}\footnote{\url{https://github.com/CEA-COSMIC/ModOpt}},  \texttt{Alpha}-\texttt{Transform}\footnote{\url{https://github.com/dedale-fet/alpha-transform}} and \texttt{Matplotlib}~\cite{hunter2007matplotlib}.

In order to study the influence of $\gamma$ from eq.~\ref{eq:loss_OSCAR}, we perform a two-step grid search by first determining the magnitude of the optimal parameter, then testing a finer grid of values in that range. The criterion chosen is
\begin{equation}
    \gamma_* = \underset{\gamma}{\text{argmin }} \delta_e (\gamma)\quad,\text{where }\delta_e(\gamma)=\underset{i}{\text{mean}}\left[\text{MSE}\left(e\left(\hat{x}_{\gamma,i}\right)\right)\right],
\end{equation}
such that $\underset{i}{\text{mean}}(x_i)$ is the mean of $\left(x_i\right)_i$ over $i$ and $\hat{x}_{\gamma,i}$ is the SCORE estimation of the $i^{th}$ galaxy with trade-off parameter equal to $\gamma$. The resulting $\gamma_*$ are shown, per $\text{SNR}$ level, in Table~\ref{tab:gamma_star}. Its value is close to 1 in all cases.

\begin{table}
\caption{Values of $\gamma_\ast$.}
\centering
\label{tab:gamma_star}
\begin{tabular}{|c|c|c|}
\hline
\rule[-1ex]{0pt}{3.5ex}  \text{SNR} & Denoising & Full restoration  \\
\hline
\rule[-1ex]{0pt}{3.5ex}  40  & 1.2 & 1.2   \\
\hline
\rule[-1ex]{0pt}{3.5ex}  75  & 0.8 & 1.6  \\
\hline
\rule[-1ex]{0pt}{3.5ex}  150 & 1.0 & 1.2  \\
\hline
\rule[-1ex]{0pt}{3.5ex}  380 & 0.8 & 0.6  \\
\hline 
\end{tabular}
\end{table}

\subsection{Results}
\label{subsec:4_2}

\subsection*{Denoising}


\begin{figure}
\vbox{\center
 \includegraphics[width= 0.95\textwidth]{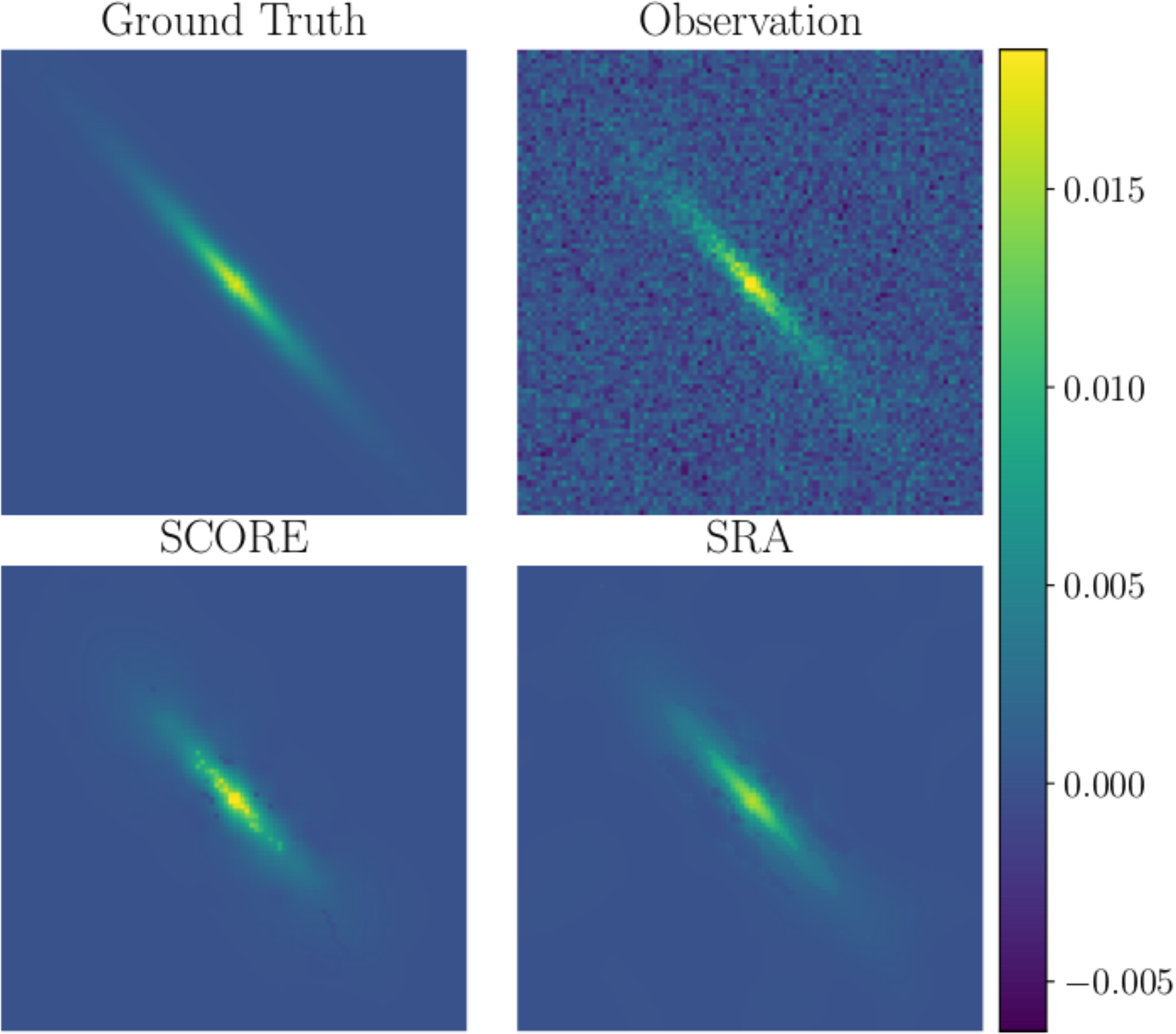} 
 \includegraphics[width= 0.95\textwidth]{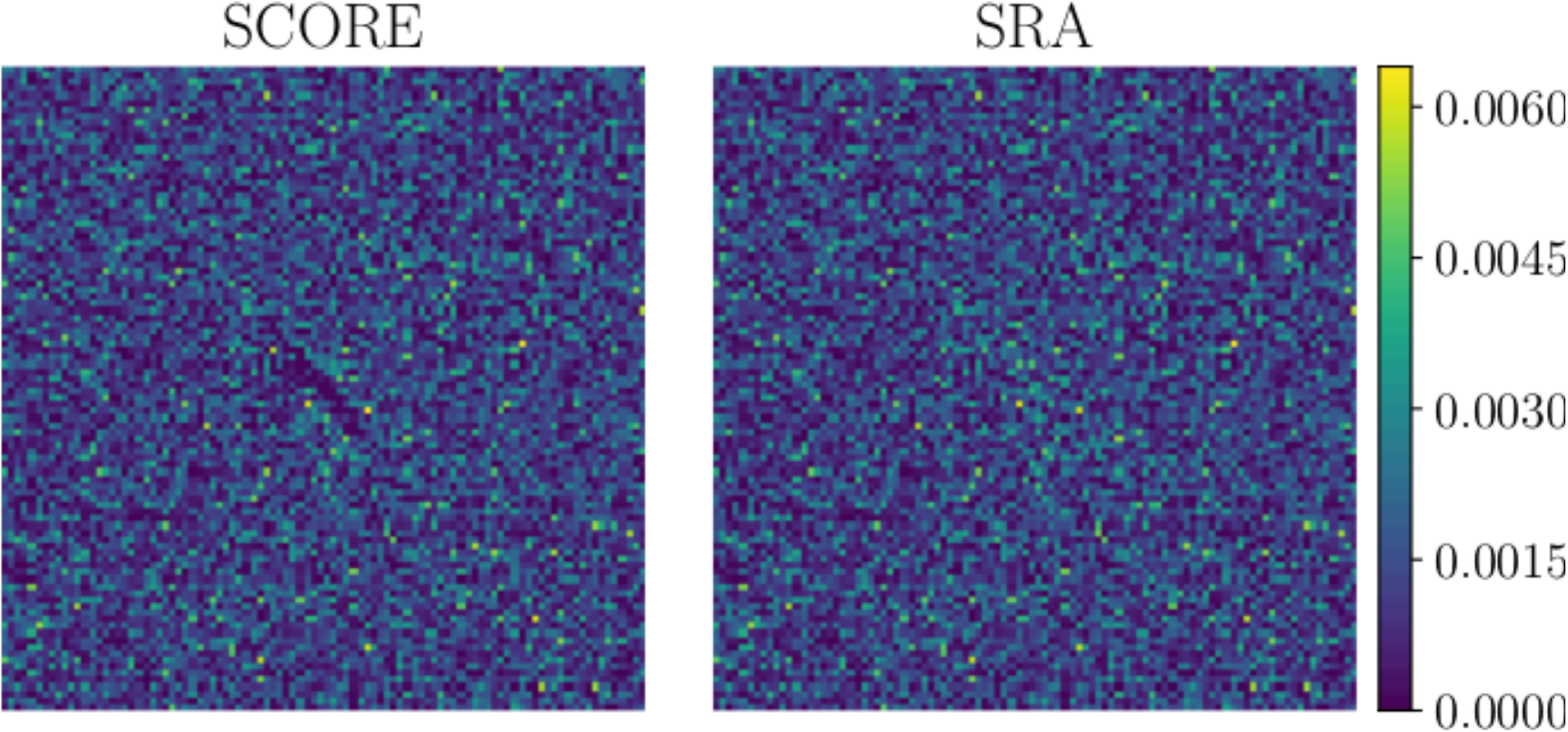}
  }
  \caption{Denoising results of galaxy \#38 for SNR=75.  Top: original image and observed data (i.e.  blurred image with noise).  Center: denoised images with SCORE and SRA. Bottom: residual images with SCORE and SRA. \href{https://github.com/CosmoStat/score/blob/master/reproducible_research/paper_results/figure_denoising.ipynb}{\faFileCodeO}}
\label{fig:den_gal38_SNR75}       
\end{figure}


\begin{figure}
\hbox{
 \includegraphics[width=0.5\textwidth]{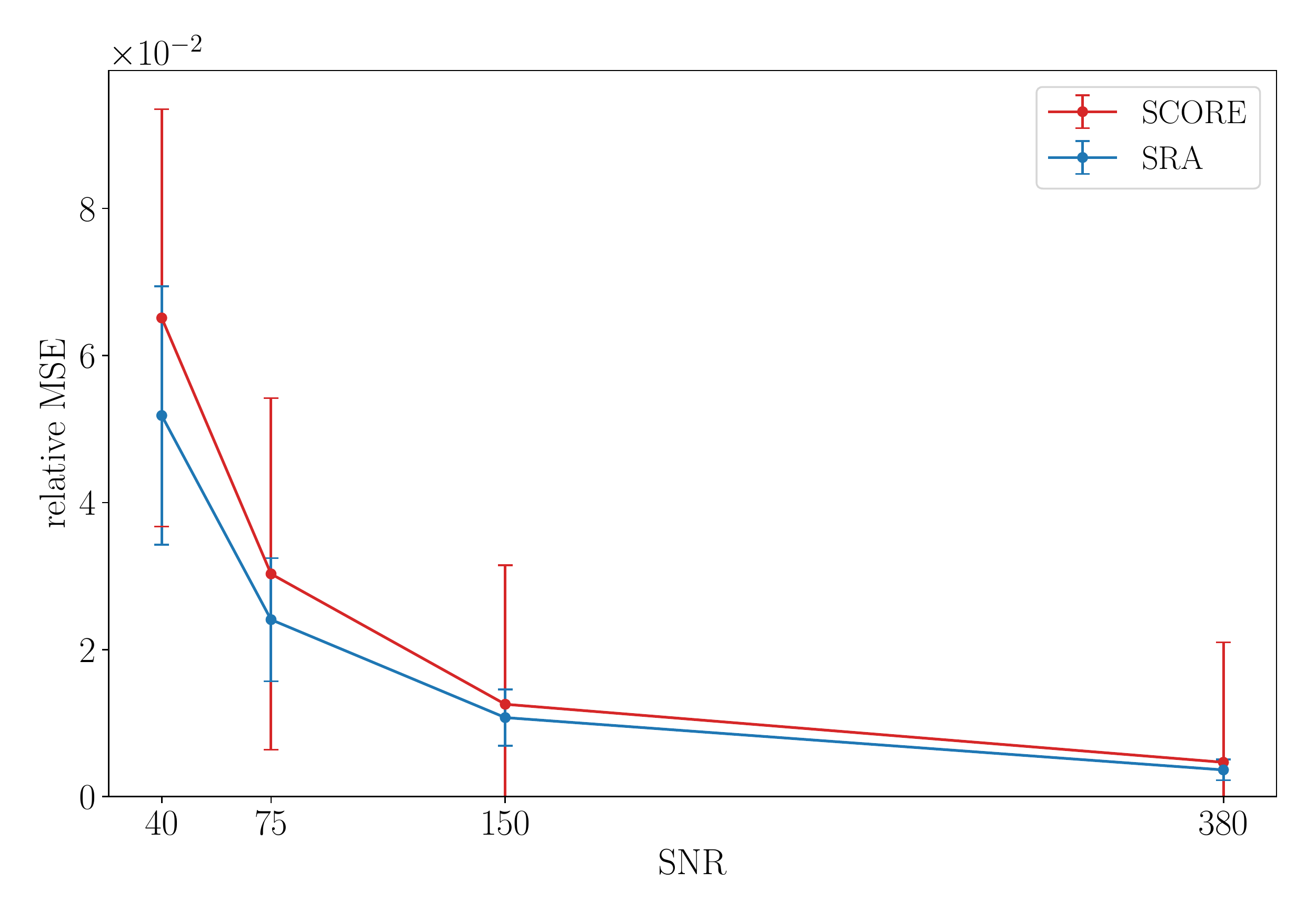}
\includegraphics[width=0.5\textwidth]{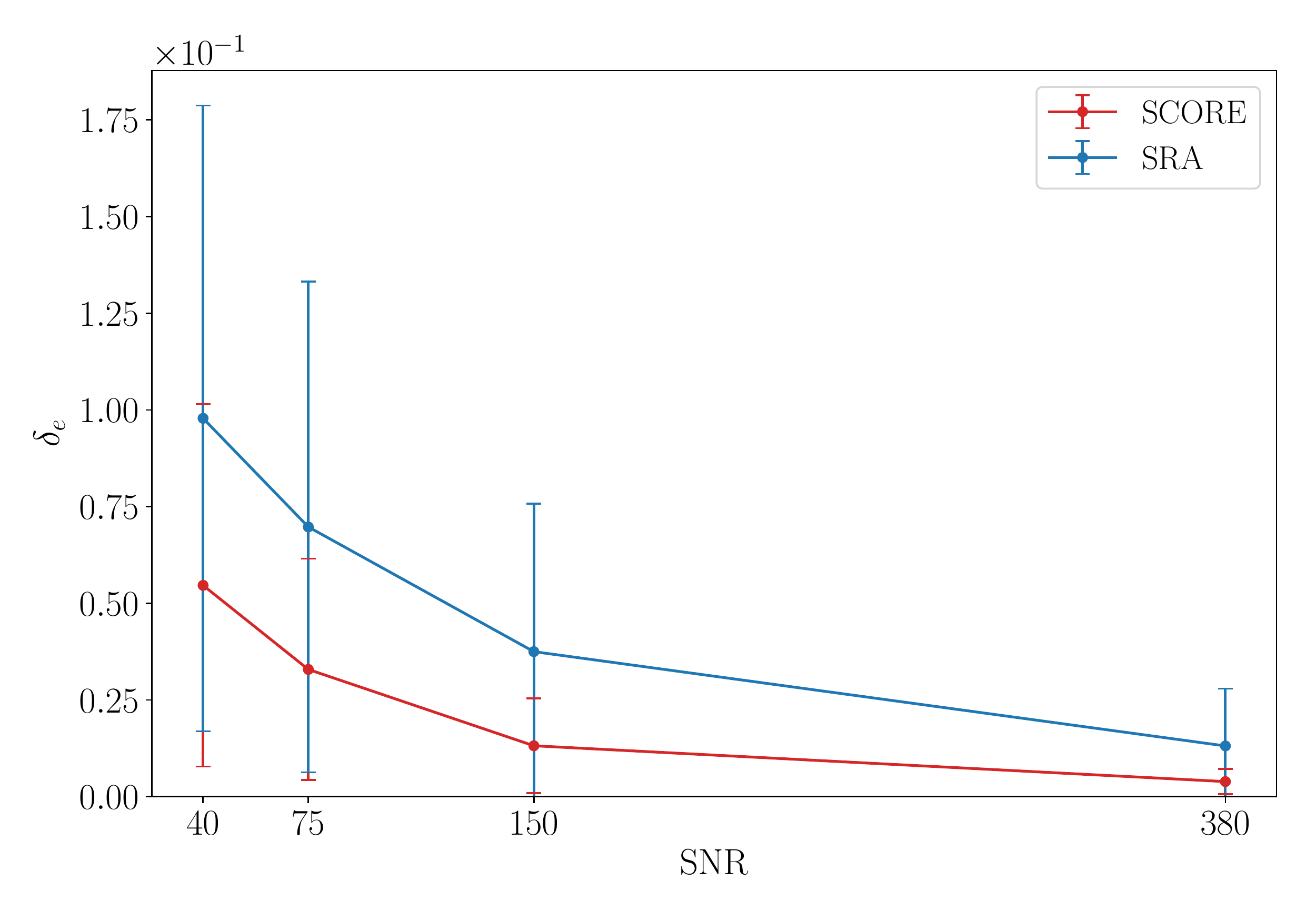}
}
\caption{Left: relative MSE per SNR of the galaxies for the denoising experiment. Right: ellipticity error $\delta_e$ per SNR. In both cases, the curves correspond to the mean per SNR and the vertical bars to the standard deviation. \href{https://github.com/CosmoStat/score/blob/master/reproducible_research/paper_results/plots_denoising.ipynb}{\faFileCodeO}}
\label{fig:denSNR_MSE}      
\end{figure}

We first consider the denoising case ($h=\delta$). 
The top row of Fig.~\ref{fig:den_gal38_SNR75} shows an example of an original galaxy image 
and  its corresponding degraded observation, the center row shows the denoised images with both SCORE and  SRA, and the bottom row the corresponding residual images.  
Fig.~\ref{fig:denSNR_MSE} shows the MSE and  ellipticity errors $\delta_e$ as a function of SNR. 

We can see that SCORE leads to a slight degradation in pixel MSE, compared to SRA. This is not unexpected as the latter's data fidelity term is entirely expressed in the image domain, while that of SCORE is shared with a shape component, as shown in sect. \ref{subsec:3_1}. SCORE's ellipticity errors are significantly reduced, by a factor of about 2.
 
 
\subsection*{Deconvolution}

\begin{figure}
\vbox{\center
 \includegraphics[width= 0.95\textwidth]{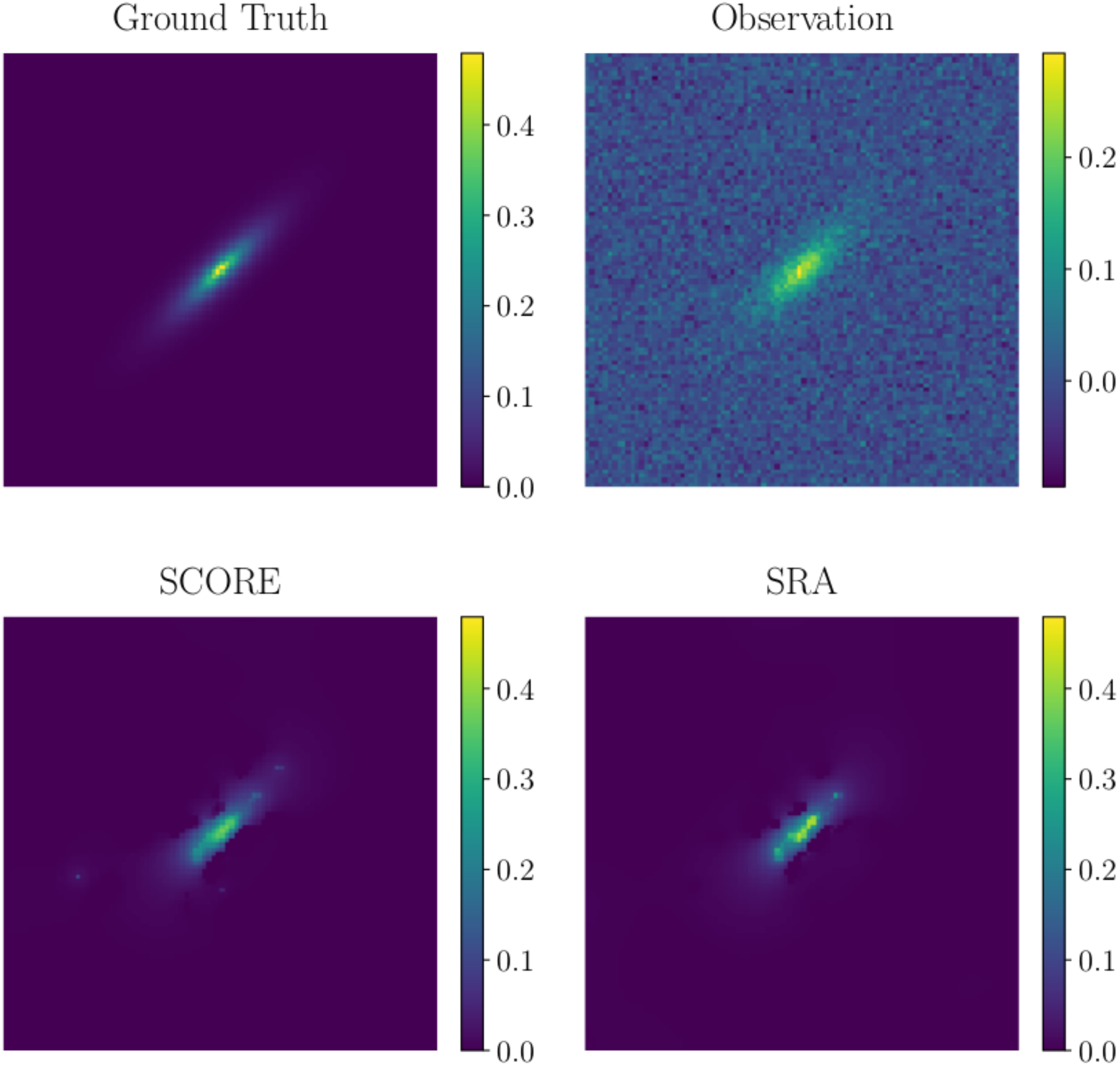}\vspace*{+3mm}  
 \includegraphics[width= 0.95\textwidth]{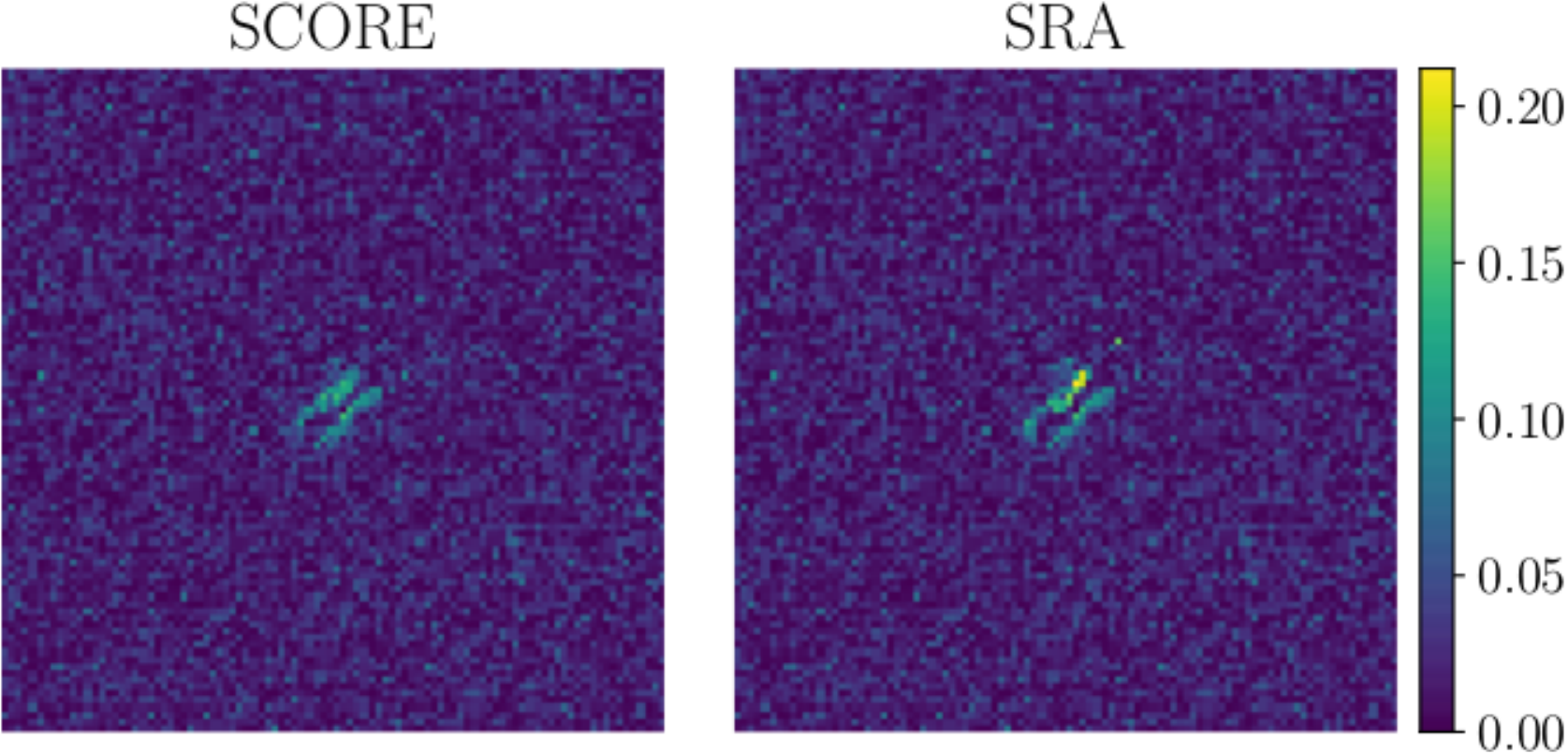}
  }
  \caption{Deconvolution results of galaxy \#16 for SNR=75.  Top: original image and observed data (i.e.  blurred image with noise).  Center: deconvolved images with SCORE and SRA. Bottom: residual images for SCORE and SRA, using the same color bar. \href{https://github.com/CosmoStat/score/blob/master/reproducible_research/paper_results/figure_deconvolution.ipynb}{\faFileCodeO}}
\label{fig:dec_gal16_SNR75}      \end{figure}

\begin{figure}
\hbox{
 \includegraphics[width=0.5\textwidth]{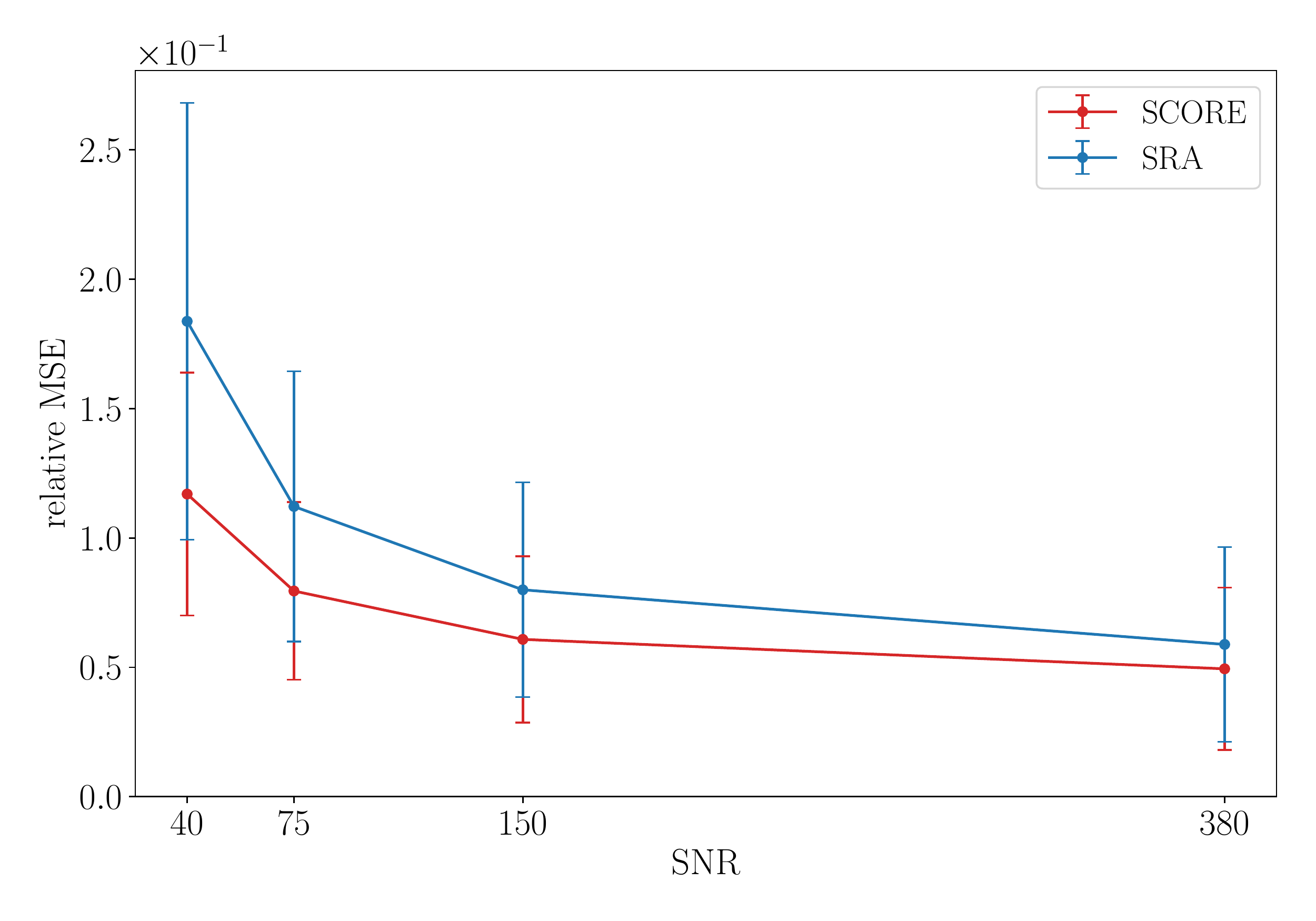}
\includegraphics[width=0.5\textwidth]{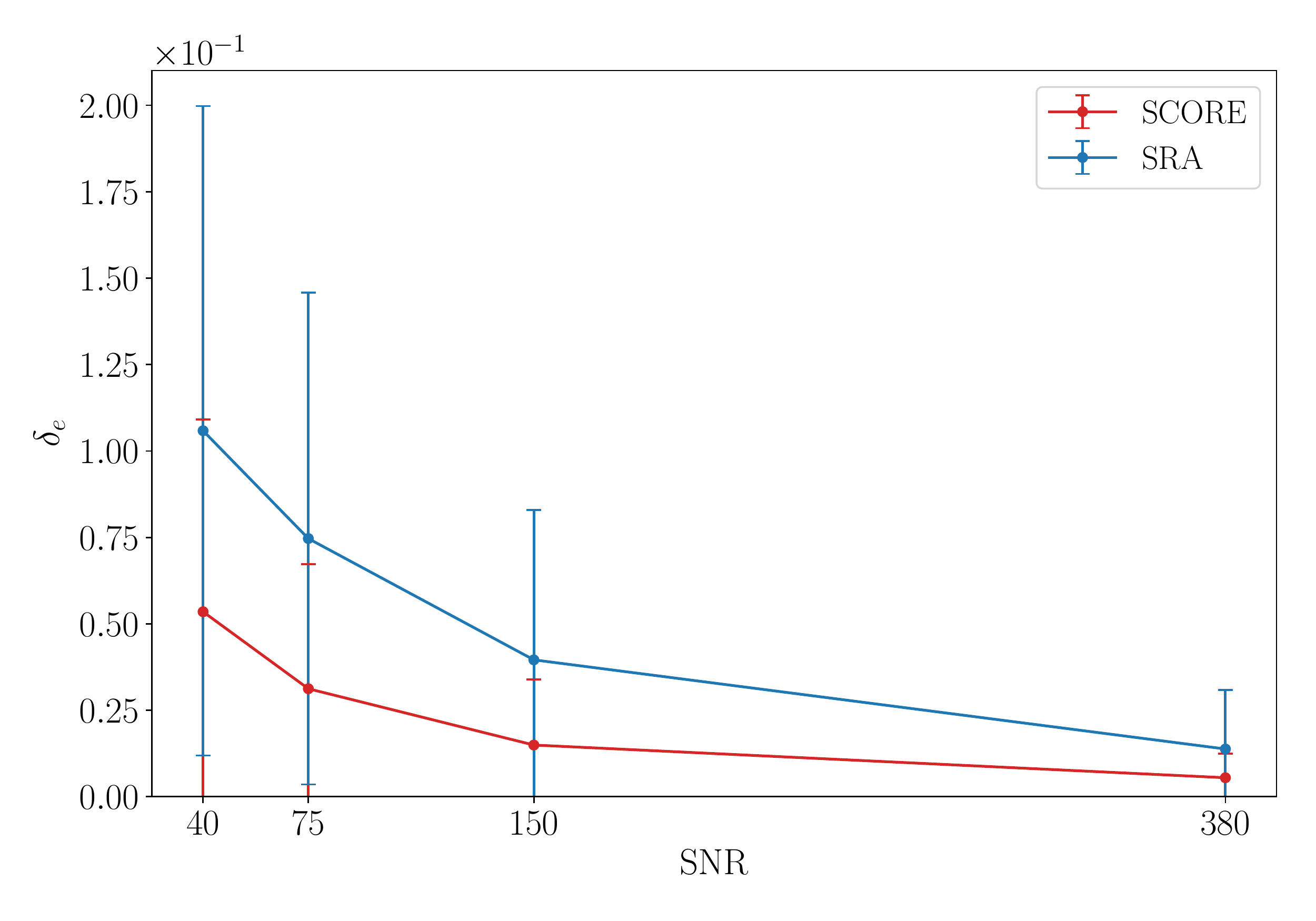}
}
\caption{Same as Fig.~\ref{fig:denSNR_MSE}, for the deconvolution experiment. \href{https://github.com/CosmoStat/score/blob/master/reproducible_research/paper_results/plots_deconvolution.ipynb}{\faFileCodeO}}
\label{fig:decSNR_MSE}      
\end{figure}

Similarly, Fig.~\ref{fig:dec_gal16_SNR75} shows an example galaxy, its recovered profiles with both approaches, and the corresponding residuals, while Fig.~\ref{fig:decSNR_MSE} shows the distributions of pixel and ellipticity errors at all SNRs.

In the case of deconvolution, SCORE performs better than SRA
for both MSE and ellipticity errors. Indeed,
the MSE yielded by SCORE is lower by at least 16\% (and 36.3\% at most) compared to SRA. The example of Fig.~\ref{fig:decSNR_MSE} illustrates that SCORE's output has a smoother profile, with a better restoration of the tail of the galaxy compared to SRA. Additionally, the residual of SCORE is, towards the center of the object, fainter than that of SRA. 

We observe different trends when looking at pixel MSE between the denoising case and the full restoration one. We believe this is due to the different conditioning of the two problems. The deconvolution  is more ill-conditioned than a simple denoising. Therefore, the broader the space of solutions, the higher the chance that an additional constraint would bring the solution closer to the ground truth. 

In terms of ellipticity, SCORE's $\delta_e$ is not only lower than SRA, but seems 
 also less biased and more consistent according to the error bars. In the denoising case, it is 44.1\% at least (and 70.3\% at most) lower, and in the deconvolution case, 49.5\% at least (and 62.3\% at most). Figs.~\ref{fig:den_gal38_SNR75} and~\ref{fig:dec_gal16_SNR75}  show that the galaxy's profile and its shape are better preserved with SCORE than with SRA.


%% file: sec5.tex
In the spirit of repeatable and reproducible research, all the codes and the resulting material have been made publicly available on GitHub at the following link: \url{https://github.com/CosmoStat/score}. In addition, at the end of the description of each figure, this icon \faFileCodeO \space provides a hyperlink to a Jupyter Notebook that shows how to generate the figure.

%% file: conclu.tex
In order to better preserve  the shapes of galaxies during a restoration process, we have proposed a new regularization term, based on the second-order moments. We have shown that our shape constraint can easily be plugged into a sparse recovery algorithm, leading to a new method called SCORE.
For denoising, when comparing to sparse recovery, 
we have shown that adding the shape constraint leads to a trade-off between the mean square error and the galaxy ellipticity error, where the latter is reduced by at least 44.1\%, while the MSE is, however, increased by at most 28.5\%. For deconvolution, both are improved 
(by at least 49.5\% for the ellipticity error and 16.9\% for the MSE). We believe this different behavior 
between denoising and deconvolution is due to the fact that the space of potential solutions is large in the deconvolution case, and that the additive shape regularizing term helps to better constrain the inverse problem.

Additionally, SCORE contains only one parameter that cannot be chosen analytically, namely the trade-off between the data-fidelity and the shape constraint terms, $\gamma$. 

We have used in this paper the Forward-Backward sparse deconvolution algorithm for its simplicity, but any more recent proximal algorithm such as Condat-Vu \cite{Condat11,Vu11} could be used as well.
The shape constraint could also easily be added to other existing deconvolution techniques, such as Total Variation \cite{weiss-tv-nesterov} or  deep learning~\cite{deepmassJ,sureau2019deep}. Finally, the shape constraint could also be improved by using moments at higher orders and also in different spaces ~\cite{flusser20162d,kostkova2019image,shakibaei2014image,wojak2010introducing}.

%% file: ack.tex
We would like to thank Christophe Kervazo, Tob\'ias Liaudat, Florent Sureau, Konstantinos Themelis, Samuel Farrens, J\'er\^ome Bobin, Ming Jiang, Axel Guinot and Jan Flusser for useful discussions. 

%% file: append_A.tex
Assume that we have $x \in \mathbb{R}^{n\times n}$ such that
\begin{equation}\label{eq:e_with_mu}
    \mathrm{e}(x) = \frac{\mu_{2,0}(x)-\mu_{0,2}(x)+2i\mu_{1,1}}{\mu_{2,0}(x)+\mu_{0,2}(x)}\quad,
\end{equation}
where $\mu_{s,t}$ is a centered moment of order $(s+t)$, defined as follows :
\begin{equation}
    \mu_{s,t}(x)=\sum_{i=1}^n\sum_{j=1}^n x\left[(i-1) n+j\right] (i-i_c)^s (j-j_c)^t \quad,
\end{equation}
and $(i_c,j_c)$ are the coordinates of the centroid of the two dimensional image encoded by $x$ :
\begin{equation}
    i_c = \frac{\sum_{i=1}^n\sum_{j=1}^n i\cdot x[(i-1) n+j]}{\sum_{i=1}^n\sum_{j=1}^n x[(i-1) n+j]} \text{ and } j_c = \frac{\sum_{i=1}^n\sum_{j=1}^n j\cdot x[(i-1) n+j]}{\sum_{i=1}^n\sum_{j=1}^n x[(i-1) n+j]} \quad.
\end{equation}

We want to show that
\begin{equation}\label{eq:e_with_u}
    \mathrm{e}(x) = \frac{\left<x,u_3\right>\left<x,u_5\right>-\left<x,u_1\right>^2+\left<x,u_2\right>^2+2i\left(\left<x,u_3\right>\left<x,u_6\right>-\left<x,u_1\right>\left<x,u_2\right>\right)}{\left<x,u_3\right>\left<x,u_4\right>-\left<x,u_1\right>^2-\left<x,u_2\right>^2} \quad,
\end{equation}
where $\forall i,j \in \{1,\dots,n\}$,
\begin{equation}
    \begin{aligned}
        &u_1[(i-1)n+j] = (i), &&u_2[(i-1)n+j] = (j),\\
        &u_3[(i-1)n+j] = (1), &&u_4[(i-1)n+j] = (i^2+j^2),\\
        &u_5[(i-1)n+j] = (i^2-j^2), &&u_6[(i-1)n+j] = (ij).
    \end{aligned}
\end{equation}
To do so, we only have to express $\mu_{0,2}(x)$; $\mu_{1,1}(x)$ and $\mu_{2,0}(x)$ using $\left(\left<x,u_i\right>\right)_{1\leq i\leq6}$. We start by introducing $m_{s,t}(x)$, the non-centered moment of order $(s+t)$,
\begin{equation}
    m_{s,t} = \sum_{i=1}^n\sum_{j=1}^n x\left[(i-1) n+j\right] i^s j^t \quad.
\end{equation}
We then express $\mu_{0,2}(x)$; $\mu_{1,1}(x)$ and $\mu_{2,0}(x)$ using the non-centered moments (of order equal or less than 2), as follows:
\begin{equation} \label{eq:mu2m}
    \begin{aligned}
        \mu_{0,2}(x) &=m_{0,2}(x)-\frac{m^2_{0,1}(x)}{m_{0,0}(x)} \quad,\\
        \mu_{1,1}(x) &=m_{1,1}(x)-\frac{m_{0,1}(x)\cdot m_{1,0}(x)}{m_{0,0}(x)}\quad,\\
        \mu_{2,0}(x) &=m_{2,0}(x)-\frac{m^2_{1,0}(x)}{m_{0,0}(x)}\quad.
    \end{aligned}
\end{equation}
Expressing these non-centered moments using $\left(\left<x,u_i\right>\right)_{1\leq i\leq6}$, we obtain
\begin{equation} \label{eq:m2u}
    \begin{aligned}
        &m_{0,0}(x) = \left<x,u_3\right>, &&m_{1,0}(x) = \left<x,u_1\right>,\\
        &m_{0,1}(x) = \left<x,u_2\right>, &&m_{1,1}(x) = \left<x,u_6\right>,\\
        &m_{0,2}(x) =\frac{1}{2}\left(\left<x,u_4\right>-\left<x,u_5\right>\right), &&m_{2,0}(x) = \frac{1}{2}\left(\left<x,u_4\right>+\left<x,u_5\right>\right).
    \end{aligned}
\end{equation}
Using eq.~\ref{eq:mu2m} and ~\ref{eq:m2u}, we can express $\mu_{0,2}(x)$, $\mu_{1,1}(x)$, and $\mu_{2,0}(x)$ using $\left(\left<x,u_i\right>\right)_{1\leq i\leq6}$:
\begin{equation} \label{eq:mu2u}
    \begin{aligned}
        \mu_{0,2}(x) &=\frac{1}{2}\left(\left<x,u_4\right>-\left<x,u_5\right>\right)-\frac{\left<x,u_2\right>^2}{\left<x,u_3\right>},\\
        \mu_{1,1}(x) &=\left<x,u_6\right>-\frac{\left<x,u_1\right>\left<x,u_2\right>}{\left<x,u_3\right>},\\
        \mu_{2,0}(x) &=\frac{1}{2}\left(\left<x,u_4\right>+\left<x,u_5\right>\right)-\frac{\left<x,u_1\right>^2}{\left<x,u_3\right>}.
    \end{aligned}
\end{equation}
We finish the proof by inserting eq.~\ref{eq:mu2u} in eq.~\ref{eq:e_with_mu} to obtain eq.~\ref{eq:e_with_u}.

%% file: append_B.tex
To generate the galaxy and the PSF images, we chose simple and commonly used profiles in astrophysics which are repectively Sersic and Moffat profiles. First, the light intensity $I_G$ of a galaxy is modeled with a Sersic using the following formula:
\begin{equation}
    I_G(R)=I_e \cdot\text{exp}\left(b_n\left[\left(\frac{R}{R_e}\right)^{\frac{1}{n}}-1\right]\right)\quad,
\end{equation}
where $n \in \mathbb{R}_+$ is the Sersic index, $R_e$ is the half-light radius, $I_e$ is the light intensity at $R_e$  and  $b_n$ satisfies $\gamma(2n;b_n)=\frac{1}{2}\Gamma(2n)$ with $\Gamma$ and $\gamma$ respectively the Gamma function and the lower incomplete gamma function. We draw the values of the parameters $n$, $I_e$ and $R_e$ from the catalog COSMOS~\cite{Mandelbaum_2011} to generate isotropic galaxy images to which we will later give a non-zero ellipticity. 

Second, the light intensity $I_P$ of a PSF is modeled with a Moffat profile using the following formula:
\begin{equation}
    I_P(R) = 2\frac{\beta-1}{\sigma^2}\left(1+\left[\frac{R}{\sigma}\right]^2\right)^{-\beta} \quad,
\end{equation}
where $\beta$ is set to 4.765 (cf. ref.~\cite{trujillo2001effects}) and $\sigma$ is calculated using the following relation:
\begin{equation}
    \text{FWHM} = 2\sigma \sqrt{2^{\frac{1}{\beta}}-1}\quad,
\end{equation}
where FWHM is the Full Width Half Maximum of the Moffat profile. Its value is drawn from a uniform distribution between 0.1 and 0.2 arcsec, which correspond respectively to Hubble Space Telescope\footnote{\url{https://www.nasa.gov/mission_pages/hubble/main/index.html}} and Euclid space telescope\footnote{\url{https://www.euclid-ec.org}} observations. This gives us preliminary, isotropic PSF images. 

We finally give an ellipticity to the both these galaxy and PSF images.  To do so, we draw the values of the ellipticity components from a centered normal distribution truncated between -1 and 1. The standard deviations are chosen as 0.3 for the galaxies and 0.03 for the PSF~\cite{bernstein2014bayesian}.